\documentclass{article}    

\usepackage{graphicx}

\title{\textbf{Schr\"{o}dinger's Cat: \mbox{The rules of engagement}}}  
\author{Richard Mould\footnote{Department of Physics and Astronomy, State University of New York, Stony Brook,
\mbox{New York} 11794-3800; http://nuclear.physics.sunysb.edu/ \~{}mould}}  
\date{}    
   
\begin{document}             

\maketitle              

\begin{abstract}

      	In a previous paper we examined the role of a conscious observer in a typical quantum mechanical
measurement.  Four rules were given that were found to govern the stochastic choice and state reduction in several
cases of continuous and intermittent observation.  It was shown that consciousness always accompanies a state
reduction leading to observation, but its presence is not sufficient to `cause' a reduction.  The distinction is
clarified and codified by the rules that are repeated below.  In this paper, these 
rules are successfully applied to two different  versions of the Schr\"{o}dinger cat experiment.   

\end{abstract}

\section*{Introduction}

In the previous paper \cite{RM}, an interaction was studied involving a particle passing over a detector
with a probability that it will either be captured, or that it will pass undetected.  A conscious observer
witnesses the detector at various times during the interaction.

It was found that when a conscious observer follows the evolution of the detector's state, consciousness always
accompanies the state reduction associated with a measurement of the particle.  During the several cases that were
investigated, consciousness was found to switch from one detector state to another on the occasion of a  stochastic
choice.   

Four rules were proposed in the previous paper that correctly describe the expected outcome in all these cases. 
The first of these rules, given below, refers to the \emph{probability current} $J$ that flows into a state.  The
current $J$ is defined to be the time rate of change of the square modulus.

\vspace{0.5 cm}

\noindent
\textbf{Rule (1)}: \emph{For any subsystem of n components in an isolated system with a square modulus equal to s, the
probability per unit time of a stochastic choice of one of those components at time t is given by $(\Sigma_nJ_n)/s$,
where the net probability current $J_n$ going into the $n^{th}$ component at that time is positive.  }

\vspace{0.5 cm}

The \emph{ready brain state} referred to in rule (2) is defined as one that is not conscious, but is physiologically
capable of becoming conscious if it is stochastically chosen.  An \emph{active} brain state is one that is either
conscious or ready.

\vspace{0.4 cm}

\noindent
\textbf{Rule (2)}: \emph{If the Hamiltonian gives rise to new components that are not classically continuous with the
old components or with each other, then all active brain states that are included in the new components will be ready
brain states.}

\vspace{0.4 cm}

\noindent
\textbf{Rule (3)}: \emph{If a component that is entangled with a ready brain state B is stochastically chosen, then B
will become conscious, and all other components will be immediately reduced to zero.}

\vspace{0.4 cm}

\noindent
\textbf{Rule (4)}: \emph{A transition between two components is forbidden if each is an entanglement containing a
ready brain state of the same observer}

\vspace{0.4 cm}

The purpose of the present paper is to apply these rules to two versions of the Schr\"{o}dinger cat experiment. 
Version I is a somewhat modified formulation of that famous puzzle.  It usually involves a cat being placed on two
components of a quantum mechanical superposition, where it is alive on one component and dead on the other.   This
distinction is ambiguous because an alive cat can be unconscious, in which case it is every bit as inert as a dead
cat.  The distinction used here is that the cat is conscious on one component of the superposition, and unconscious
on the other.  In version II, the cat begins in an unconscious state, and is aroused to a conscious state.

\section*{The Apparatus}

The apparatus will consist of a radioactive source and detector that will be denoted by either $D_0$ or $D_1$, where
the first means that the detector has not yet received a decay particle, and the second means that it has.  The
detector output will be connected to a mechanical device that carries out a certain task, such as a hammer falling on a
container that then  releases an anesthetic gas.  This device will be denoted by $M(\alpha, t)$, where $\alpha$
indicates the extent to which the task has been completed, and $t$ is the time.  $D_0M(\alpha_0, t)$ means that the
source has not yet decayed at time $t$ and that the mechanical device is still in its initial position given by
$\alpha_0$. $D_1M(\alpha_1, t)$ means that the decay has already occurred by the time $t$, and that the mechanical
device has  advanced to a position given by $\alpha_1$.  Let $I_0$ be an indicator that tells us that $M$ has
not completed its task, and $I_1$ tells us that it has.  Then $D_1M(\alpha_1, t)I_0$ means that the device has not
completed its task at time $t$.  When $\alpha = \alpha_f$, we will say that the device $M$ has fully run its course,
so $D_1M(\alpha_f, t)I_1$ means that the source has decayed, and that the mechanical device has completed its task at
time $t$ as indicated by $\alpha_f$ and by the indicator $I_1$.  We also suppose that the source is exposed to the
detector for a time that is limited to the half-life of a single emission.  At that time a clock will shut off the
detector, so it will remain in the state $D_0$ if there has not yet been a particle capture. 

The system at $t_0$ = 0 is then:  $\Phi(t_0) = D_0M(\alpha_0, t_0)I_0$, and in time it becomes 
\begin{equation}
\Phi(t\ge t_0) = D_0M(\alpha_0, t)I_0 + \int{dz}D_1M(\alpha_z, t)I_0 + D_1M(\alpha_f, t)I_1 
\end{equation}
where   $\int{dz}D_1M(\alpha_z, t)I_0$ and $D_1M(\alpha_f, t)I_1$ are zero at $t_0$, and $z$ covers the range $0  
\le z < f$.   The significance of the integral is that at some time $t$ there is a spread of alphas that
represent the possible state of the mechanical device at that moment.  Although the device is a classical object,
there is a quantum mechanical uncertainty as to when it begins its operation.  The function $M(\alpha_z, t)$ is
therefore a pulse that represents that uncertainty moving along the $+ z$ axis.  As time progresses, the second
component in eq.\ 1 and then the third component will gain in amplitude, but the third component cannot do that until
after a time
$T$ that corresponds to the time it takes for the mechanical device to complete its task.  	

Since we arranged to have the first component decrease for a time equal to the half-life of a single emission, its
square modulus will stabilize to a constant value of 0.5 at that time, assuming that eq.\ 1 is normalized.  After that,
no new current will flow into the second component, so its amplitude will fall back to zero as the pulse in
$M(\alpha_z, t)$  runs out along $+ z$.  When $M(\alpha_z, t)$ finally goes to zero, the third component will reach
its maximum value.  In the end, the first and third components will survive, each with a square modulus equal to 0.5.

\section*{Sequential Interactions with an Observer}

This apparatus involves two sequential interactions: the radioactive decay and the operation of the
mechanical device.  The previous paper (ref.\ 1) did not consider more than one interaction, so before inflicting this
apparatus on a cat, we will see how the rules work when an outside observer witnesses the apparatus in operation.   

Let the external observer look at the apparatus at some time $t_{ob}$ after the process has begun.  Equation 1 will
then become 
\begin{eqnarray}
\lefteqn{\Phi(t\ge t_{ob}) =}\\
& & \hspace{.38cm}D_0M(\alpha_0, t)I_0X\ + \int{dz}D_1M(\alpha_z,t)I_0X + D_1M(\alpha_f, t)I_1X   \nonumber\\
& & + \hspace{.1cm} D_0M'(\alpha_0, t)I_0B_0 + \int{dz}D_1M'(\alpha_z,t)I_0B_1 + D_1M'(\alpha_f, t)I_1B_1 \nonumber
\end{eqnarray}
where $X$ is the unknown brain state of the observer prior to the observation, and $B_0$ and $B_1$ are normalized ready
brain states of the observer that perceive $D_0$ and $D_1$ respectively.  The three ready brain components are zero at
$t_{ob}$, and thereafter receive current from the corresponding component in the first row of eq.\ 2.  That current is
due to the physiological interaction that occurs when the observer interacts (e.g., visually) with the apparatus.      

Rule (1) with $n = 3$ (treating the integral as one component) requires that the time integrated current flowing into
the second row of eq.\ 2 must equal 1.0.  So one or the other component must be eventually chosen.  We take them in
reverse order.

If the observation occurs after time $T$, current will have flowed into the last component $D_1M'(\alpha_f, t)I_1B_1$
of eq.\ 2.  If that component happens to be stochastically chosen at a time $t_{scf}$, then the ready state $B_1$ will
become conscious, and following rule (3), 
\begin{displaymath}
\Phi(t\ge t_{scf} > t_{ob}) = D_1M(\alpha_f)I_1\underline{B}_1
\end{displaymath}
This will complete the interaction.  It corresponds to the observer coming on the scene when the mechanical device has
already finished its task.  As in the previous paper, underlining a brain state such as $\underline{B}_1$ means that it
is a conscious state.

The fifth component in eq.\ 2 containing the integral is a continuum of components in the variable $\alpha_z$ at time
$t$.   If one of those components is stochastically chosen at a time $t_{sc1}$,  the corresponding value
$\alpha_{sc1}$ will be selected at that time.  That choice will make $B_1$ conscious, and following rule (3), the
system will become 
\begin{equation}
\Phi(t = t_{sc1} > t_{ob}) = D_1M(\alpha_{sc1})I_0\underline{B}_1
\end{equation}

From this point on the observer will track the behavior of the mechanical device like a classical observer.  The
Hamiltonian will carry the mechanical device through its paces from $\alpha_{sc1}$ to $\alpha_f$, while the conscious
observer $\underline{B}_1$ remains focused on the detector state $D_1$.

Another possibility is that there will be a stochastic choice of the fourth component in eq.\ 2 at a time $t_{sc0}$. 
According to rule (1), this can only happen while that component is still increasing; and that can only happen before
the radioactive source has reached the single emission half-life time $t_{1/2}$, inasmuch as the detector is cut off
at that time.  Assuming that this time has not run out, and that the fourth component is stochastically chosen,
then $B_0$ will become conscious, giving
\begin{displaymath}
\Phi(t_{1/2}> t = t_{sc0}> t_{ob}) = D_0M(\alpha_0)I_0\underline{B}_0
\end{displaymath}
This corresponds to the outside observer coming upon the apparatus \emph{before} the radioactive source has decayed. 
The system will then continue to evolve, starting at the new time, 
\begin{equation}
\Phi(t_{1/2}> t \ge t_{sc0} > t_{ob}) = D_0M(\alpha_0, t)I_0\underline{B}_0 + D_1M''(\alpha_0, t)I_0{B}_1
\end{equation}
where $D_1M''(\alpha_0, t)I_0B_1$ is zero at $t_{sc0}$. This component will not take the form of an integral
over $\alpha_z$ because the Hamiltonian will only connect the first component with the second component in eq.\ 4; and
in addition, rule (4) will not allow a self-generating succession of ready brain states.  That is, a
transition to a component $\alpha_z$ containing a ready brain state $B_1$ is not allowed if it can only get there from
another component $\alpha_{z'}$ containing  the ready brain state $B_1$. Consequently, the component
$\alpha_0$ cannot be skipped over as the mechanical device begins its operation.  The significance of this is
discussed in the last paragraph of this section.   

If the second component in eq.\ 4 is stochastically chosen at
time $t_{sc1'}$ such that  $t_{1/2} > t_{sc1'} > t_{sc0}$, then the system will again be reduced, giving 
\begin{displaymath}
\Phi(t_{1/2} > t = t_{sc1'} > t_{sc0} > t_{ob}) = D_1M(\alpha_0)I_0\underline{B}_1
\end{displaymath}
From this point on, the observer will track the classical behavior of the mechanical device as happened following
eq.\ 3.  In this case it begins with $\alpha_0$.

And finally, if the fourth component of eq.\ 2 is stochastically chosen but the second component of eq.\ 4 is
\emph{not} chosen, then the first component (of eq.\ 4) will run out the half-life time on the clock, rendering
its output current equal to zero.  When that happens  eq.\ 4 will stabilize to give
\begin{equation}
\Phi(t \ge t_{1/2}) = D_0M(\alpha_0)I_0\underline{B}_0 +  D_1M''(\alpha_0)I_0{B}_1
\end{equation}
where each component has come to the same constant square modulus.  The continuing existence of this \emph{residual
superposition} is not unphysical.  It is like similar cases in the previous paper (ref.\ 1) where the conscious
observer on one component is unaware of the other (not conscious) component.  It was shown in that paper that the
rules require another reduction if a second observer looks in on the scene in eq.\ 5, or if the conscious attention of
the primary observer is allowed to drift in a non-classical way. This reduction will eliminate the second component in
\mbox{eq. 5}.  We call this a ``phantom" component because it servers no further purpose at this point.  Equation 5
therefore corresponds to the observer finding the detector in the state $D_0$ with the clock run out. 

The clock limiting the detector is set to equal the half-life of a single emission, and this means that there is a 50\%
chance that the system will be given by \mbox{eq.\ 5}.  That will happen if a stochastic choice of the fourth component
of \mbox{eq.\ 2} is chosen at time $t_{sc0}$; and if subsequently, the stochastic choice of the second component in
eq.\ 4 is not chosen at time $t_{sc1'}$.  Otherwise, there is a 50\% chance that the outside observer will witness
the mechanical device complete its operation to the end. 

It should be noted that rule (4) saves us from another anomaly that is different in kind from the one noted in the
previous paper.  If the second component of eq.\ 4 were an integral over $\alpha_z$, eq.\ 5 would never be able to
stabilize as a residual superposition.  That's because a pulse of ready brain states would then run through the second
component, and its leading edge would be continuously receiving current from the trailing edge. The pulse is not
normalized, but since it keeps using the same current over and over again, rule (1) guarantees that there will
eventually be a stochastic hit on a ready brain component within the pulse.  That  guarantees a reduction in which
the first component in \mbox{eq.\ 5} will become zero.  This is an anomalous result because it would prevent
the observer from ever finding the detector in the state $D_0$ with the clock run out.

\section*{Version I}

We now replace the indicator in eq.\ 1 with a cat that is initially in the conscious state $\underline{C}_0$ as shown
in the first component of eq.\ 6.  The mechanical devise is one that will render the cat unconscious, as represented by
the state $U$ in the last component of eq.\ 6.  As in the previous paper, we require that the lower physiological
functions leading to $U$ are included in the variable $\alpha$ of the mechanical device, just prior to its reaching
$\alpha_f$.  This means that the mechanical device in eq.\ 6 is different to this extent from the device in  eq.\ 1. 
Before a stochastic choice occurs, the system would then \emph{apparently} be given by 
\begin{eqnarray}
\Phi( t \ge t_0) = D_0M(\alpha_0, t)\underline{C}_0  +\int{dz}D_1M(\alpha_z, t)C_1 + D_1M(\alpha_f, t)U
\end{eqnarray}
where the last two components are initially equal to zero, and where $C_1$ is a ready bran state of the cat.  However,
the conditions of the experiment require that the cat is still conscious when the mechanical device begins its operation
at
$\alpha_0$; so
$\alpha$ in the second component in \mbox{eq.\ 6} must have a sub-zero value when that component becomes active.  Here
again,
$\alpha_0$ cannot be skipped over.  This means that the last component and all but the first component under the
integral are not really present in   eq.\ 6, since they cannot possibility appear before there has been a stochastic
choice.  This requirement is enforced by rule (4) that again forbids a self-generated integral of ready brain states.
Equation 6  should therefore be written
\begin{equation}
\Phi(t\ge t_0) = D_0M(\alpha_0, t)\underline{C}_0 + D_1M'(\alpha_0, t)C_1
\end{equation}
where $D_1M'(\alpha_0, t)C_1$ is zero at $t_0$.  If this component is stochastically chosen at time $t_{sc}$, then the
system will be 
\begin{equation}
\Phi(t_{sc}) =  D_1M(\alpha_0)\underline{C}_1
\end{equation}
From this point on, the cat will track the classical behavior of the mechanical device until it completes the task of
rendering the creature unconscious at time $t_f$.  The final state of the cat is then given by
\begin{equation}
\Phi(t \ge t_f) =  D_1M(\alpha_f)U
\end{equation}

If, on the other hand, the second component in eq.\ 7 is not stochastically chosen, then the components will stop
interacting at the half-life time $t_{1/2}$, so current will cease flowing from the first to the second component in
that equation.  This means that eq.\ 7 will stabilize in place giving 
\begin{equation}
\Phi(t \ge t_{1/2}) =  D_0M(\alpha_0)\underline{C}_0 + D_1M'(\alpha_0)C_1
\end{equation}
where both components have come to a square modulus equal to 0.5, assuming that eq.\ 7 is initially normalized.  As in
eq.\ 5, the cat will be conscious without any awareness of the other component, so the phantom component of this
residual superposition does no harm.   

There is a 50\% possibility that eq.\ 9 will be the final state, and a 50\% possibility that eq.\ 10 will be the final
state.  This outcome conforms to normal expectations, so rules (1) - (4) are adequate to the task.

\section*{Paradox Lost}

Equation 10 shows a conscious cat on one component of the superposition with a  square modulus equal to 0.5, 
and a non-conscious cat on the other component with the same square modulus.  This is a form that is generally said to
be paradoxical.  How, it is asked, can the cat have the same `intrinsic' probability of being both conscious and
non-conscious at the same time?  The question suggests that the cat's state is truly enigmatic. But that is not so.  

In the first place, it is not correct in this treatment to say that either component has an `intrinsic' probability of
any kind.  Probability is associated only with current flow, not with the magnitude of a square modulus.  There
is no current flow in eq.\ 10.  Second, the cat is unambiguously conscious in this superposition.  The cat would
certainly say so, and so would an outside observer for whom the second component is only a phantom.   
There is therefore nothing paradoxical about eq.\ 10.

\section*{Version I with Outside Observer}

Imagine that an outside observer looks in on the cat \emph{during} these proceedings to see how it is doing.  If that
happens after the cat has engaged the mechanical device in the classical progression following eq.\ 8, then the
observer and the cat will together follow the classical working out of the mechanical device.

If the outside observer interacts with the system before a stochastic choice causes the cat to become classically
engaged, then eq.\ 7 becomes
\begin{eqnarray}
\Phi(t \ge t_{ob} > t_0) &=&  D_0M(\alpha_0, t)\underline{C}_0X + D_1M'(\alpha_0, t)C_1X \\
 &+& D_0M''(\alpha_0,t)\underline{C}_0B_0\nonumber
\end{eqnarray}
where the third component is equal to zero at $t_{ob}$.  A fourth component is forbidden by rule (4).    

If the second component of eq.\ 11 is stochastically chosen, the result will be the same as eq.\ 8 with the outside
observer still ``outside".  This corresponds to the case in which the mechanical device begins its operation after the
observation but before the second observer can (physiologically) come on board.  Of course, he will be on board as
soon as his physiological processes permit, and from that point on he will follow the classical evolution of the cat. 
   
	If the third component in eq.\ 11 is selected at time $t_{sc0}$, this will correspond to the conscious observer
joining the conscious cat before the mechanical interaction has begun.  The result would
be $D_0M(\alpha_0)\underline{C}_0\underline{B}_0$, and its continuing evolution would yield 
\begin{displaymath}
\Phi(t \ge t_{sc0}) = D_0M(\alpha_0, t)\underline{C}_0\underline{B}_0 + D_1M''(\alpha_0,
t)C_1B_1
\end{displaymath}
where the second component is equal to zero at $t = t_{sc0}$.  This is the same as \mbox{eq.\ 7}, except that the
second observer is now on board with the cat and will follow its classical fate in eqs.\ 8 and 9, or join it in the
residual superposition of eq.\ 10.  

The total probability is found from rule (1) with $n = 2$ involving  integrals of current into the second and third
components of eq.\ 11 that are taken from $t_0$ to the end of both the mechanical and physiological interactions.
\begin{displaymath}
\int [J_x + J_0]dt = 1
\end{displaymath}
where $J_x$ and $J_0$ go into $D_1M'(\alpha_0, t)C_1X$ and $\hspace{.03cm} D_0M''(\alpha_0, t)\underline{C}_0B_0$. 
So if the second component of eq.\ 11 is not stochastically chosen,  it is certain that the third component will be
chosen.

\section*{Version II}

In the second version of the Schr\"{o}dinger cat experiment, the cat is initially unconscious, and is awakened by an
alarm that is set off by the capture of a radioactive decay particle.  The mechanical device $M(\alpha, t)$ is now an
alarm clock, where $\alpha$ represents the successive stages that progress from the initial ring to the low level
physiological processes that terminate in the cat's ready brain state.  As before, the alarm will only go off 50\% of
the time.
\begin{equation}
\Phi(t\ge t_0) = D_0M(\alpha_0, t)U + \int{dz}D_1M(\alpha_z, t)U + D_1M(\alpha_f, t)C
\end{equation}
where $U$ is the unconscious state of the cat, $C$ is the cat's ready brain state, and the second and third components
are initially equal to zero.  Variable $z$ covers the range $0 \le z < f$.   Again, there may be a time delay $T$
before the third component containing the ready brain state of the cat can accumulate value after $t_0$.  We assume
eq.\ 12 to be normalized.

	When current does flow into the third component it might be stochastically chosen at time $t_{sc}$.  If that happens,
then the system will become
\begin{equation}
\Phi(t\ge t_{sc}) = D_1M(\alpha_f)\underline{C}
\end{equation}
This will terminate the interaction.  It corresponds to the cat finding himself aroused by the alarm, and this will
happen 50 \% of the time.

	Only the third component in eq.\ 12 contains a ready brain state, so only it can be stochastically chosen in a way
that leads to a rule (3) reduction.  If there is no stochastic choice, then the square modulus of the first component
of eq.\ 12 will fall to a value of 0.5.  The second component will initially rise to some positive value
and fall again to zero, and the third component will rise to a square modulus of 0.5.  In the final state of the
system, the square moduli of the first and third components will be equal to 0.5, and the second component will be
zero.  Therefore, some time $t_f$ after the alarm mechanism has run its course, the system will end its evolution in
the superposition 
\begin{equation}
\Phi(t> t_f) = D_0M(\alpha_0)U + D_1M(\alpha_f)C
\end{equation}
which will appear 50\% of the time.  This superposition will only be reduced if there is an outside observer, or if the
cat wakes up naturally.  We will take these two cases separately

\section*{Version II with Outside Observer}

	If the outside observer makes contact with the cat \& apparatus after there has been a stochastic choice leading to
eq.\ 13, then following a separate physiological interaction, the conscious observer will be on board with the
conscious cat.  The two of them will then experience an amended version of eq.\ 13
given by 
\begin{equation}
\Phi(t \ge t_{sc}) = D_1M(\alpha_f)\underline{C}\underline{B}_1
\end{equation}
where $\underline{B}_1$ is the conscious state of the outside observer.   

	Now imagine that the outside observer enters the picture before the stochastic choice that leads to eq.\ 13.  Equation
12 would then become
\begin{eqnarray}
\lefteqn{\Phi(t_{sc} >t \ge t_{ob}) =}\\
& & \hspace{.39cm}D_0M(\alpha_0, t)UX + \int{dz}D_1M(\alpha_z,t)UX + D_1M(\alpha_f, t)CX \nonumber\\
& &  +\hspace{.1cm} D_0M'(\alpha_0, t)UB_0 + \int{dz}D_1M'(\alpha_z,t)UB_1 \nonumber
\end{eqnarray}
where the primed components in the second row are equal to zero at $t_{ob}$.  A sixth component is not allowed by rule
(4).  

	If the third component in eq.\ 16 is stochastically chosen, realizing the component  $D_1M(\alpha_f,
t)\underline{C}X$, then the continuing physiological interaction will bring about a transition from $X$ to $B_1$, which
will result in a final state $D_1M(\alpha_f)\underline{CB}_1$ like eq.\ 15.  This corresponds to the cat becoming
conscious after the observation but before the observer has had time to climb on board. 

	If $M(\alpha_{sc1})$ in the fifth component in eq.\ 16 is stochastically chosen at the time $t_{sc1}$, this will
result in the state $D_1M(\alpha_{sc1})U\underline{B}_1$.  The outside observer will then be on board with the
unconscious cat when the mechanical device has reached the stage given by $\alpha_{sc1}$.  From that point on, the
observer will track the classical operation of the alarm prior to its awakening the cat.  This, in turn, leads to a
final state that also adds to eq.\ 15.  

	If the fourth component in eq.\ 16 is stochastically chosen at time $t_{sc0}$, then we will have the state
$D_0M(\alpha_{0})U\underline{B}_0$.  This will happen if the observer intervenes prior to the time that a radioactive
particle is captured by the detector.   In that case, the decay interaction will begin again giving 
\begin{equation}
\Phi(t_{1/2} >t \ge t_{sc0} > t_{ob}) = D_0M(\alpha_0, t)U\underline{B}_0 + D_1M''(\alpha_0, t)UB_1
\end{equation}
where the second component is zero at time $t_{sc0}$.  Rule (4) forbids the second component from generating ready
brain components that are successors to $\alpha_0$, so the only transition that is possible from the first component
 is one going to the $\alpha_0$ component.  Again, $\alpha_0$ cannot be passed over.

If there is a subsequent stochastic choice at time $t_{sc1'}$, then the state in eq.\ 17 will become
$D_1M(\alpha_0)U\underline{B}_1$, and the observer will classically track the slumbering cat from the time the alarm
mechanism is first launched to the end.  This too will lead to a final state that adds to eq.\ 15.  

	The final possibility is that there will be no stochastic choice at $t_{sc1'}$, in which case the first term in eq.\
17 will stabilize at the half-life time $t_{1/2}$.  When that happens, we will have
\begin{equation}
\Phi(t > t_{1/2}) = D_0M(\alpha_0)U\underline{B}_0 + D_1M''(\alpha_0)UB_1
\end{equation}
where the square modulus of each of the components is equal to 0.5.  As in previous cases, the residual
superposition in eq.\ 18 will be reduced if the outside observer's consciousness drifts away from $\underline{B}_0$, or
if another outside observer looks in on the experiment.  Since the second component is a phantom, eq.\ 18 corresponds
to the observer finding the cat unconscious when the clock has run out.  This happens 50 \% of the time, and eq.\ 15
happens 50 \% of the time.

\section*{Version II with a Natural Wake-Up}

	Even if the alarm does not go off, the cat will wake up naturally by virtue of its own internal alarm clock.  The
internal alarm can be represented by a classical mechanical device that operates during the same time as the external
alarm.  The interaction is assumed to run parallel to eq.\ 12, and is given by
\begin{displaymath}
\int{dz}N(\beta_z, t)U + N(\beta_f, t)C_N
\end{displaymath}
where $N(\beta)$ is the internal mechanism in the variable $\beta$, and $\beta_f$ is the final value that accompanies
the associated ready body state $C_N$ of the cat.  The integral covers the range $0 \le z < f$, where $\beta_0$ is
the initial value of $\beta$.  As with the external alarm, the internal mechanism takes a time $T_N$ to complete its
task, so $N(\beta_f, t)C_N$ will follow from  $\int{dz}N(\beta_z, t)U$ only after that time has elapsed.  Because
it is classical, the z-wave running through the integral will be very sharply defined.

When the experiment begins, both the internal and external mechanisms will run  parallel to one another, starting at
the same time $t_0$.

\begin{eqnarray}
\Phi(t\ge t_0) &=& D_0M(\alpha_0, t)  \{\int{dz}N(\beta_z, t)U + N(\beta_f, t)C_N\} \\
&+& \{\int{dz'}D_1M(\alpha_{z'}, t)  \}\{\int{dz}N(\beta_z, t)U +  N(\beta_f, t)C_N\}\nonumber \\
&+& D_1M(\alpha_f, t)  \{\int{dz}N(\beta_z, t)C \} \nonumber
\end{eqnarray}
where $0\le z, z' <f$, the first component $D_0M(a_0, t) \int{dz}N(\beta_z, t)U$ is normalized to 1.0 at $t_0$, and
all of the other components are initially zero.  As before, $C$ represents the ready body state of the cat that can be
aroused by the external alarm.   The sixth component in eq.\ 19 does not exist because  $\alpha_f$ and
$\beta_f$ are contradictory body states, so they are not permitted to emerge in a single component.    

As the system evolves after $t_0$, probability current will flow into the components containing $C_N$ and $C$, creating
the possibility that one of them will be stochastically chosen.  Since both are terminal states that arise from a
single source state, rule (1) requires that the probability that one of them is chosen in time $dt$ is equal to
$(J_{C_N} + J_C)dt$.  Integrating this from $t_0$ to the end of both interactions at time $t_{ff}$ gives a total
probability of 1.0 that one of the body states will be stochastically chosen.  

We stipulated in eq.\ 14 that the cat was not awakened by the external alarm, even after the middle component in that
equation had fallen off to zero.  Applying this condition to eq.\ 19 gives   
\pagebreak
\begin{eqnarray}
\Phi(t\ge t_0) &=& D_0M(\alpha_0, t)  \{\int{dz}N(\beta_z, t)U + N(\beta_f, t)C_N\}\\
&+& D_1M(\alpha_f, t)\int{dz}N(\beta_z, t)C \nonumber
\end{eqnarray}
Under these circumstances, the probability of a stochastic choice of either $C$ or $C_N$ would be 0.5.   Since we
stated as a condition that $C$ is not chosen, it is certain that $C_N$ will be chosen by the time both interactions are
complete. When that happens, eq.\ 20 becomes
\begin{displaymath}
\Phi(t > t_{ff}) =D_0M(\alpha_0) N(\beta_f)\underline{C}_N
\end{displaymath}
This corresponds to the cat waking up naturally to find that the detector has not captured a radioactive particle, and
that the mechanical device is still in its initial $\alpha_0$ position.  Therefore, the sleeping cat's internal alarm
clock does the job that the external alarm has failed to do.  When we dropped the second row in eq.\ 19, we exclude the
possibility that the cat would wake up during the operation of the mechanical device.

\end{document}